# Thermodynamics of Substances with Negative Thermal Expansion Coefficient

I. A. Stepanov

Latvian University, Rainis bulv. 19, Riga, LV-1586, Latvia

**Abstract**

The 1st law of thermodynamics for heat exchange is

$$dQ = dU + PdV.$$

According to K. Martinas etc., J. Non-Equil. Thermod. **23** (4), 351-375 (1988), for substances with negative thermal expansion coefficient, P in this law is negative. In the present paper it has been shown that P for such substances is positive but the sign before P must be minus not plus:

$$dQ = dU - PdV.$$

**Introduction**

The 1st law of thermodynamics for heat exchange has the following form:

$$TdS = dU + PdV. \qquad (1)$$

There are substances which contract with the temperature: water at T<277 K, liquid He[4], Si in a certain temperature interval, honey, Te, Se (monoclinic), quartz glass etc [1, 2].

Water contracts with the temperature at 273<T<277 K. Let's carry out the process (1) with dU=0 in this temperature interval: dQ>0 and dV<0. (In [3] is demonstrated a possibility of full conversion of heat to mechanical work: "This example provides the clearest evidence that the 2nd law of thermodynamics is untenable. There are no fundamental constraints for the conversion of heat into work, apart from the law of conservation of energy"). Throughout this paper pressure is supposed to be the atmospheric one. Then dQ=PdV. It is a contradiction. Both sides of the equation have different signs. In this paper this paradox is explained.

**Theory**

It is possible to prove that at 273<T<277 K and dU=0, water contracts with heating. From thermodynamics [1, 4]

$$\alpha=(1/V)(\partial V/\partial T)_P=-(1/V)(\partial V/\partial P)_T(\partial P/\partial T)_V=k_T(\partial S/\partial V)_T. \quad (2)$$

where $k_T$ is the isothermal compressibility. If $\alpha<0$ then $(\partial S/\partial V)_T<0$ and $(\partial S/\partial V)_U<0$ because $(\partial S/\partial V)_U$ is a special case of $(\partial S/\partial V)_T$. From (1)

$$(\partial S/\partial V)_U=P/T>0. \quad (3)$$

Really, let's suppose that water expands at U=const, 273<T=$T_0$<277 K. Then, due to continuity of the function V, there will be an interval $T_0$-ΔT<T<$T_0$+ΔT where water expands at T≠const. But it contradicts to experiment because $\lim_{\Delta T \to 0} (1/V)[(V(T_0+\Delta T)-V(T_0))/\Delta T]_P = \alpha < 0$.

Even if it is impossible to carry out a process with dU=0, one has to pay attention that chemical reaction never goes to the end, i.e. reactants never turn to products completely. Nevertheless, it is possible to calculate the whole heat of reaction exact, it means, the heat of reaction where all reactants turn to the products (enthalpy $\Delta H^0$). It is done by the Van't-Hoff equation, it is derived from the 1st and the 2nd law of thermodynamics without simplifications [5]:

$$d/dT \ln K = \Delta H^0/RT^2 \qquad (4)$$

where K is the reaction equilibrium constant.

Thermodynamics must give opportunity to calculate the change of the volume ΔV for the process with dU=0 even if this process is impossible. For substances with negative thermal expansion coefficient, equation dQ=dU+PdV will give wrong result.

To explain this paradox, in [6] it has been supposed that P in (1) is negative for substances with negative α. The author thinks that this opinion is wrong. According to foundations of thermodynamics, pressure P in (1) is the external pressure [4], here the positive atmospheric one.

According to thermodynamics,

$$P = -(\partial U/\partial V)_S. \qquad (5)$$

and P is negative for metastable states [4, 6]. In [6] it is supposed that substances with negative thermal expansion coefficient are metastable. It is doubtful. The metastable states correspond to local minima, the stable one corresponds to the absolute minimum [4]. Water at T<277 K has only one state.

For water in weightlessness P in (5) is greater than 0, regardless of is T<277 K or T>277 K. For water in cylinder under piston under pressure P=$-(\partial U/\partial V)_{S,+0}<0$, P=$-(\partial U/\partial V)_{S,-0}>0$, regardless of the temperature. Therefore, water at T<277 K does not differ from that at T>277 K in that sense. For liquid with negative pressure, P in (5) is less than zero [6] (regardless of is the derivative forward one or backward one).

In solids $P_{+0}$ in (5) is negative, however, it does not mean that the sign before P in (1) must be negative. For almost all solids $\alpha>0$. Backward derivative $P_{-0}>0$.

One has to arrive at a conclusion that for substances with negative $\alpha$ the sign before P in (1) must be minus not plus:

$$dQ=dU-PdV. \qquad (6)$$

Whence $dQ=dQ_1+dQ_2$, $dQ_1=dU$, $dQ_2=\pm PdV$. If $dQ>0$, then $dQ_1>0$ and $dQ_2>0$.

Almost nobody checked experimentally validity of (1) for substances with $\alpha<0$. There are relations using which one can verify the 1st law [4, 7]: the Mayer's relation

$$c_V-c_P=-T\alpha^2 V/k_T. \qquad (7)$$

The Reech's relation

$$c_V = c_P k_S / k_T. \qquad (8)$$

$k_S = -(1/V)(\partial V/\partial P)_S$ is the adiabatic compressibility, and the relation

$$(\partial c_P/\partial P)_T = -TV((\partial \alpha/\partial T)_P + \alpha^2). \qquad (9)$$

If to derive these relations using (6) instead of (1), they will be the following ones

$$c_V - c_P = T\alpha^2 V/k_T. \qquad (10)$$

$$c_V = c_P k_S / k_T. \qquad (11)$$

$$(\partial c_P/\partial P)_T = TV((\partial \alpha/\partial T)_P + \alpha^2). \qquad (12)$$

So, one can make an important conclusion: for $\alpha < 0$, $k_S > k_T$.

For water at 273K $c_V - c_P \approx -2,5$ J/(kg·K), $c_P = 4217,6$ J/(kg·K) [7]. Using (7) one can not verify the 1st law because one does not know $c_V$. One can do it using (8): for T=273K $k_T = 5,0885 \cdot 10^{-10}$ Pa$^{-1}$=$k_S$ [7]. These values can prove neither (1) nor (6).

The authors of [7] used table dependence of enthalpy H(P, T) from a reference book to obtain $(\partial c_P/\partial P)_T = \partial^2 H/\partial P \partial T$. However, in reference books [8, 9] enthalpy is obtained from the relations

$$H = \Delta H_f^0 + \int c_P dT, \qquad (13)$$

$$(\partial H/\partial P)_T = V - T(\partial V/\partial T)_P \qquad (14)$$

where $\Delta H_f^0$ is the enthalpy of formation of substance and (9) is a sequence of (14). Hence, one may not agree with the result from [7] that $(\partial c_P/\partial P)_T > 0$ in the whole temperature range.

**Conclusions**

The following conclusion can be drawn. The author thinks he succeeded to show that (6) must be used instead of (1) for substances with negative thermal expansion coefficient. The final proof can be done by measuring precisely $c_v$, $k_T$ and $k_S$.

**References**

1. John S. O. Evans, J. Chem. Soc., Dalton Trans., 1999, (19), 3317.

2. C. A. Angell, R. D. Bressel, M. Hemmati, E. J. Sare and J. C. Tucker, PCCP, 2000, 2, 1559.

3. G. V. Skornyakov, Pis'ma Zh. Tekh. Fiz., **21**(23) (1995) 1-5, [English translation: Tech. Phys. Lett., **21**(12) (1995) 949-950].

4. Physical Encyclopaedia, Sovetskaja Entsiklopedia, Moscow, 1990.

5. F. Daniels and R. Alberty, Physical Chemistry (John Wiley & Sons, New York, 4th edition, 1975).

6. K. Martinas, A. Imre and L. P. N. Rebelo, J. Non-Equilib. Thermod., 1988, 23, 351.

7. J. Guemez, C. Fiolhais and M. Fiolhais, Am. J. Phys., 1999, 67, 1100.

8. L. V. Gurvich and I. V. Veitz, Thermodynamic Properties of Individual Substances, Hemisphere Pub Co., NY, L., 1989, etc. 4th ed.

9. S. L. Rivkin, A. A. Aleksandrov and E. A. Kremenevskaja, Thermodynamic derivatives for water and water vapour, Energija, Moscow, 1977, pp. 3-6.